\documentclass[a4paper]{jpconf} 
\usepackage{graphicx}
\usepackage{lineno}
\begin{document}
\title{Scintillation light detection in the long-drift ProtoDUNE-DP liquid argon TPC}

\author{C. Cuesta on behalf of the DUNE Collaboration}

\address{Centro de Investigaciones Energ\'{e}ticas, Medioambientales y Tecnol\'{o}gias (CIEMAT), \\ 28040 Madrid, Spain}

\ead{clara.cuesta@ciemat.es}

\begin{abstract}
ProtoDUNE-DP is a 6\,$\times$\,6\,$\times$\,6\,m$^3$ liquid argon time-projection-chamber (LArTPC) operated at the Neutrino Platform at CERN in 2019-2020 as a prototype of the DUNE Far Detector. DUNE is a dual-site experiment for long-baseline neutrino oscillation studies, neutrino astrophysics and nucleon decay searches. The light signal in these detectors is crucial to provide precise timing capabilities. In ProtoDUNE-DP, scintillation light produced by cosmic muons in the LArTPC is collected by the photomultiplier tubes (PMTs) placed up to 7 m away from the point of interaction. The scintillation light production and propagation processes are analyzed and compared to simulations, improving the understanding of some liquid argon properties.
\end{abstract}

\section{ProtoDUNE Dual Phase detector}
The Deep Underground Neutrino Experiment (DUNE) aims to address key questions in neutrino physics and astroparticle physics~\cite{DUNE_LBL, DUNE_SNB, DUNE_BSM}. DUNE will consist of a near detector placed at Fermilab close to the production point of the muon neutrino beam of the Long-Baseline Neutrino Facility (LBNF), and four 17\,kt liquid argon time-projection chambers (LArTPCs) as the far detector in the Sanford Underground Research Facility (SURF) at 1300\,km from Fermilab~\cite{DUNEtdrv4}. 

The ProtoDUNE Dual Phase (DP) detector~\cite{wa105,Cuesta:2019yeh} operated at the CERN Neutrino Platform in 2019-2020 to demonstrate the LArTPC DP technology at large scale as a possibility for the DUNE far detector. ProtoDUNE-DP has an active volume of 6$\times$6$\times$6 m$^{3}$ corresponding to a total LAr mass of 750\,t, being the largest DP LArTPC ever operated. In ProtoDUNE-DP the argon ionization creates electrons which drift vertically thanks to an electric field. The ionization charge is then extracted, amplified, and detected in gaseous argon above the liquid surface, allowing a good signal to noise ratio and a fine spatial resolution. The scintillation light signal is collected by a photon detection system (PDS) to provide a trigger, and to determine precisely the event time, with possibility to perform calorimetric measurements and particle identification. Two Cosmic Ray Tagger (CRT) planes were added to opposite walls of the ProtoDUNE-DP cryostat to trigger on muon-tracks passing through both CRTs. 

The PDS of ProtoDUNE-DP~\cite{protoDUNElight} is formed of 36 8-inch cryogenic PMTs, R5912-02MOD from Hamamatsu~\cite{protoDUNEPMTs, Belver:2020qmf}, placed below the cathode grid. As the PMTs are not sensitive to 127-nm light, a wavelength shifter is placed to convert this light to the PMT wavelength detection range. A light calibration system (LCS) is installed to obtain an equalized PMT response~\cite{Belver:2019lqm}. A dedicated light acquisition and calibration software was developed for ProtoDUNE-DP~\cite{ProtoDUNEDP_LACS}.

\section{ProtoDUNE-DP photon detection system performance}
\label{sec:per}

ProtoDUNE-DP collected cosmic-ray data for 18 months, from June 2019 until November 2020 in different conditions of electric fields. A total of 130.7 million events were acquired with a duration of 675 hours.

All 36 PMTs were operational since the beginning of the data taking and the basic performance of the PDS system is validated. A time accuracy among the channels better than 16\,ns is measured. The low noise in the baseline of the signals, $0.6\pm0.1$\,ADC, is remarkable as the baseline presents very small fluctuation and was stable with time. At a gain of $10^7$, the single photo-electron (SPE) amplitude is 7\,$\pm$\,2\,ADC counts, implying a signal-to-noise ratio greater than 11. 
 
The main goal of the LCS is to calibrate the PMT response by determining the PMT gain during the operation of the detector to measure the collected light charge in photo-electron (PE) units. The gain calibration method, based on measuring the SPE charge at a given voltage, is described in~\cite{Belver:2019lqm}. During operation, PMTs are biased at the HV required to achieve the target gain according to the calibration results. It has to be noted that the PMTs are switched on and off every day (sometimes several times on the same day). Despite this, PMT gains were quite stable with time, being $9\%$ the standard deviation of the gain at 1500\,V for all the data taken during operation for the 36 PMTs, . 
 

The LAr scintillation light is produced at 127\,nm, a wavelength for which most photosensors are not sensitive, and fluorescent materials are introduced to shift the photon wavelength towards the visible range. ProtoDUNE-DP performs this task through a mixed system of 30 PMTs covered with polyethylene naphthalate (PEN) foils, and 6 PMTs directly coated with tetraphenyl butadiene (TPB). While TPB is broadly used, PEN is a novel material, never used before in such a large scale experiment and which efficiency is not well known. The PEN sample used in ProtoDUNE-DP is transparent and biaxially oriented. It has been installed as round foils of 240\,mm diameter and 0.125\,mm thickness placed tangent to the PMT glass surface. TPB was deposited over the PMT polished surface, using a dedicated evaporation system developed by the ICARUS experiment~\cite{Bonesini_2018}.

The relative photon detection efficiency of the PEN PMTs versus the TPB PMTs is experimentally determined by the light charge detected (S1 signal). The average charge collected on the PMTs for these events is around 200\,PEs on TPB PMTs, and 50\,PEs on PEN PMTs. This means that on average, TPB PMTs detect four times more photons than PEN PMTs. Additionally, a simple model is proposed to compute the relative WLS efficiency of both materials taking into account the geometrical differences between both systems. As a result, it is estimated that TPB will produce three times more visible photons than PEN, for the same amount of incident VUV photons. 

\begin{figure}[ht]
\begin{minipage}{0.48\textwidth}
    \includegraphics[width=\textwidth]{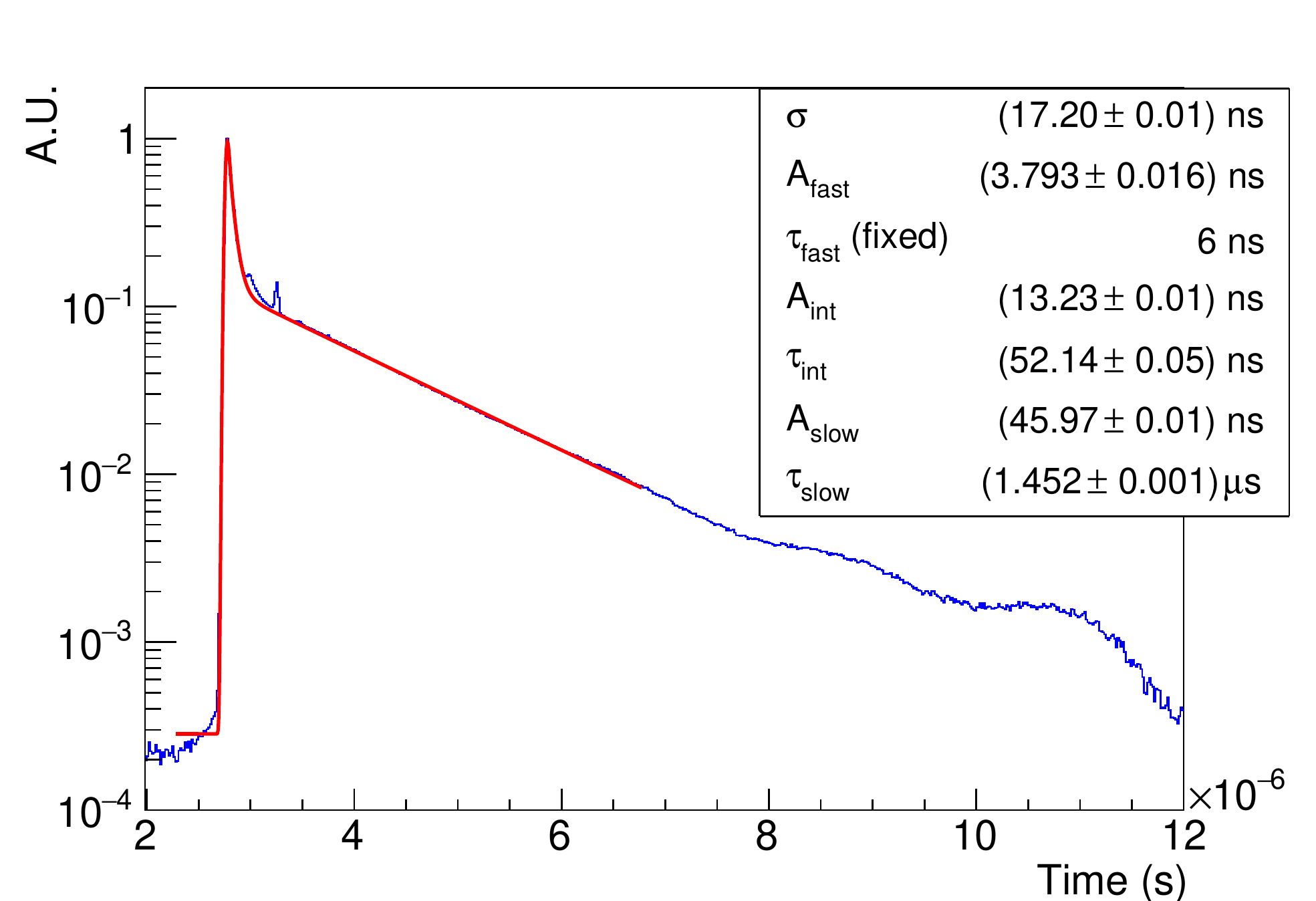}
\caption{Average scintillation waveform of a ProtoDUNE-DP PMT in blue and fitted function to the parameters described in the text in red.}\label{fig:Fit1}
\end{minipage}\hspace{0.04\textwidth}
\begin{minipage}{0.48\textwidth}
 \includegraphics[width=\textwidth]{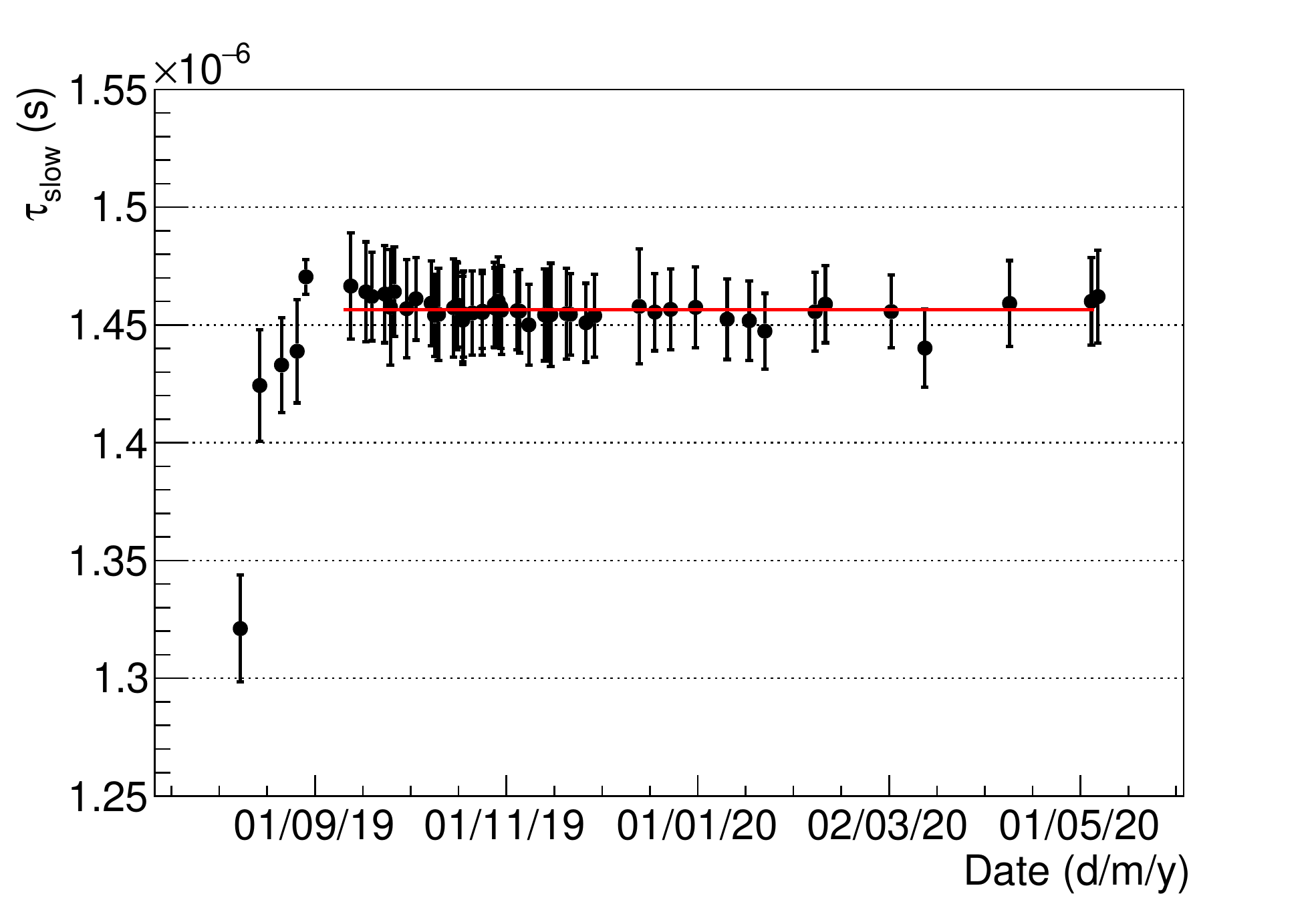}
\caption{Monitoring of the $\tau_{slow}$ component of the scintillation time profile during the ProtoDUNE-DP operation and linear fit shown in red.}\label{fig:Fit2}
\end{minipage} 
\end{figure}

The scintillation light emission in LAr, called S1 signal, has a characteristic time dependence. Waveforms are well described as a sum of three exponential functions convoluted with a Gaussian function to represent the detector response. The scintillation time profile should have two components, from the decay to ground state of singlet ($\tau_{fast}$) and triplet ($\tau_{slow}$) argon excimers, but an intermediate component ($\tau_{int}$) is added in order to improve the convergence of the fit as reported also by other LAr experiments~\cite{311light}. Figure~\ref{fig:Fit1} shows the  average waveform of muon-like events in the absence of drift field for one PMT and Fig.~\ref{fig:Fit2} the evolution of the average $\tau_{slow}$ during the operation of ProtoDUNE-DP. The value of the $\tau_{slow}=1.46 \pm 0.02$\,$\mu$s has a small dispersion over time and indicates a high LAr purity at the ppb level.

\section{Light production and propagation in LAr}
\label{sec:light}

The study of the scintillation light production, propagation and collection in a LArTPC is performed with data acquired with the CRT-trigger system to profit from the off-line reconstruction of the track trajectory. Analyses are mainly based on the correlation between the light charge and the distance from the muon track to the PMT. 

\begin{figure}[ht]
\begin{minipage}{0.45\textwidth}
   \includegraphics[width=\textwidth]{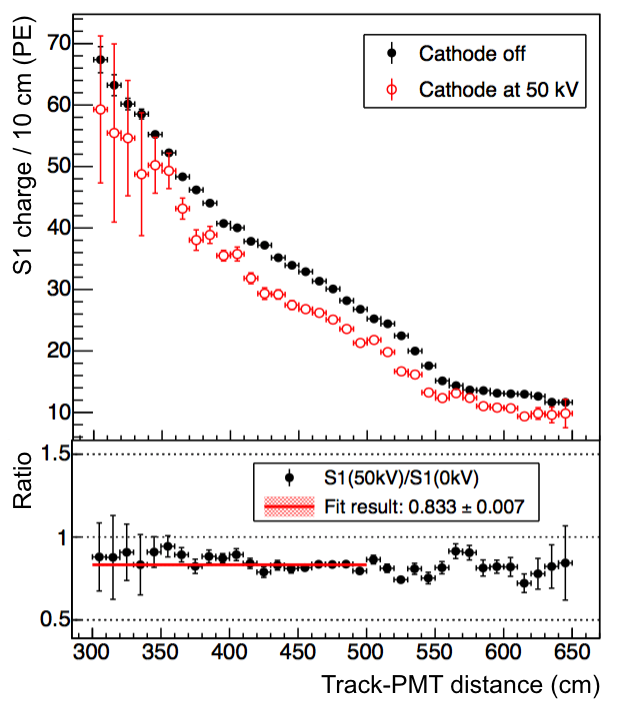}
\caption{Top panel: Average charge collected by the ProtoDUNE-DP PEN PMTs as a function of the track-PMT distance with null drift field (black) and with cathode at 50\,kV (red). Bottom panel: Ratio between the top panel distributions.}\label{fig:Fields1}
\end{minipage}\hspace{0.04\textwidth}
\begin{minipage}{0.48\textwidth}
    \includegraphics[width=\textwidth]{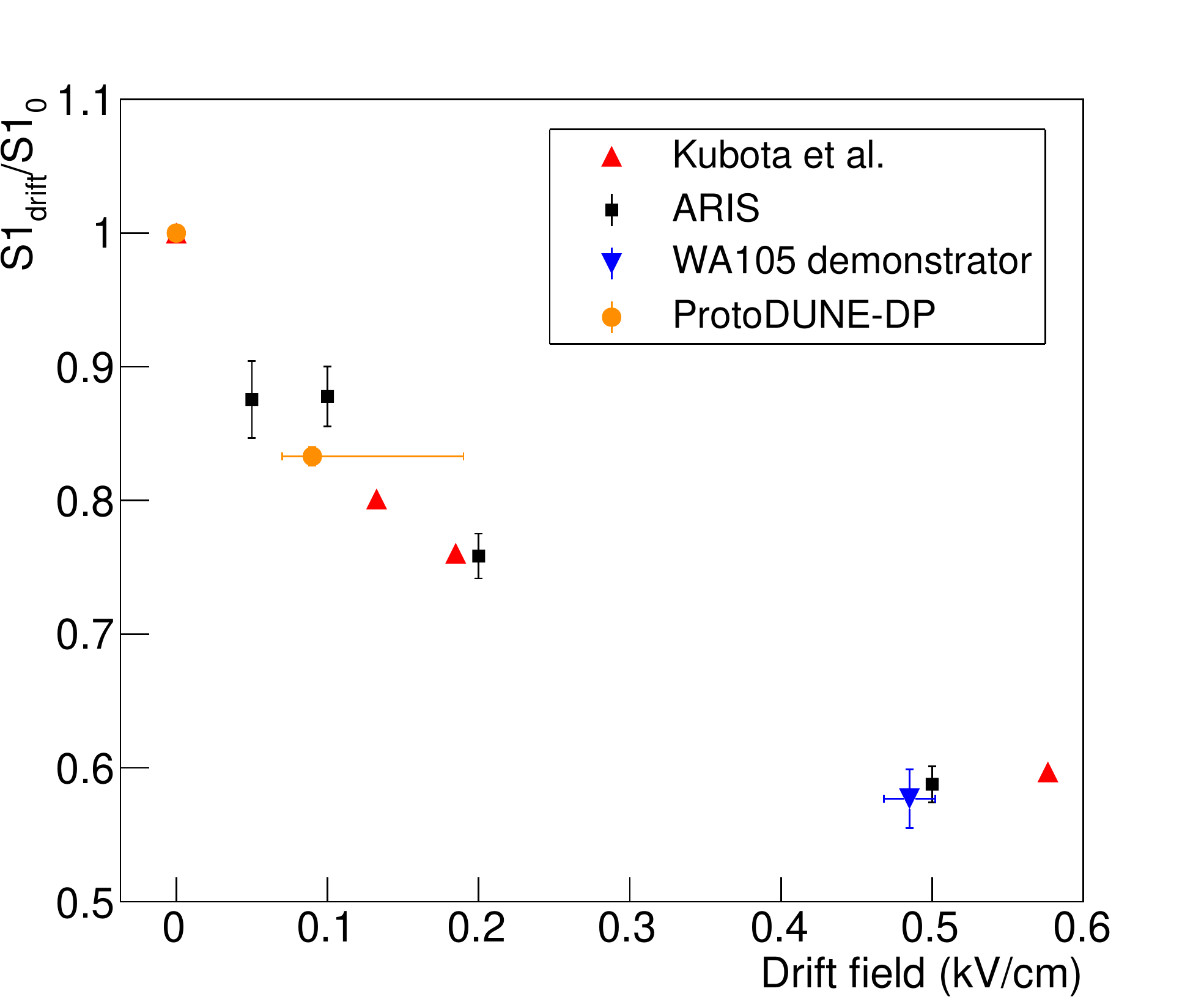}
\caption{Light yield reduction with drift field measured by different experiments~\cite{311light,Kubota,Aris}.}\label{fig:Fields2}
\end{minipage} 
\end{figure}

The suppression of the electron-ion recombination process due to the drift field implies the reduction of the primary scintillation light production. The ProtoDUNE-DP operation faced issues that impacted the TPC field conditions, as a short circuit limited the cathode HV to 50\,kV and the drift field was reduced to the top part of the drift (along $\sim$1\,m). Even so, the light levels detected with the PDS at null drift field and at 50\,kV are compared to roughly quantify such a light yield decrease in Fig.~\ref{fig:Fields1}. It is obtained that, at least, 17\% of the scintillation light detected in absence of drift field comes from electron-ion recombination.

An approximate average field is associated to the light-yield ratio to verify the empirical Birks’ law, see Fig.~\ref{fig:Fields2}. Despite the relatively large uncertainty of the ProtoDUNE-DP result, a fair agreement is found with previous works~\cite{311light,Kubota,Aris}.



The size of ProtoDUNE-DP, the longest drift-distance LArTPC ever operated, allows for an unprecedented study of the light propagation. The Rayleigh scattering length (RSL) has a quantitative impact on the amount of light collected, so an evaluation of the RSL value is carried out by comparing the measured light signals with the light predicted by the MC simulation testing two lengths (61.0\,cm and 99.9\,cm). The light attenuation modeling with an exponential function in the distance parameter allows the measurement of the overall attenuation in data and MC, and so the evaluation of the agreement for the different simulated configurations.

\begin{figure}[ht]
\begin{minipage}{\textwidth}
\centering
    \includegraphics[width=0.45\textwidth]{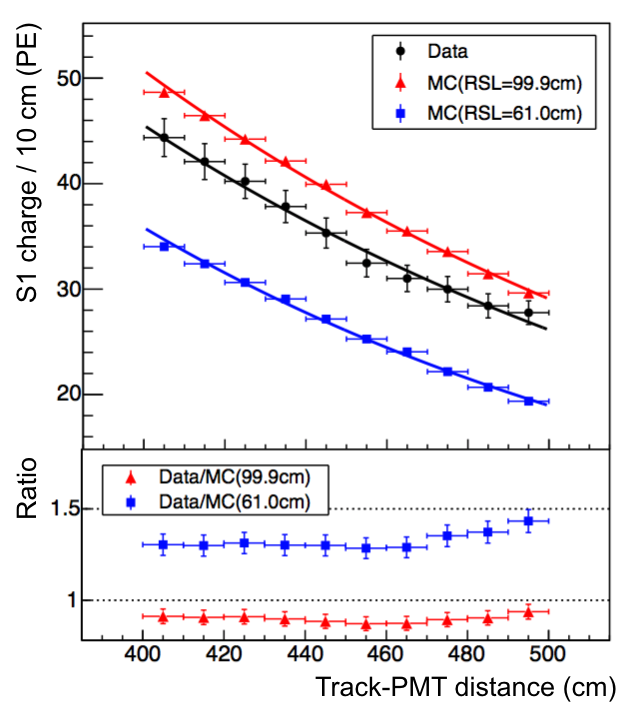}
\caption{Top panel: Average collected S1 charge by the ProtoDUNE-DP PEN PMTs as a function of the track-PMT distance. Two MC samples with different RSL values (61.0\,cm and 99.9\,cm) are compared to data. For each distribution, an exponential fit is performed (see results in Table~\ref{tab:RSL2}) and plotted as a solid line. Bottom panel: Data-MC ratio for the two previous MC samples.}\label{fig:RSL1}
\end{minipage} 
\end{figure}

In Fig.~\ref{fig:RSL1}, the light charge as a function of the distance from the muon track to the PMT is fitted to an exponential decay and the data-MC ratios for the two simulations are also presented. Looking at the distribution shape, the agreement between data and the 99.9-cm MC sample is better than with the 61.0-cm value. The attenuation length values obtained are presented in Table~\ref{tab:RSL2}, and it is observed that the data sample also agrees better with the 99.9-cm MC sample. The measured attenuation length is higher than the RSL, so the light is expected to undergo Rayleigh scattering before being deeply attenuated due to, for example, absorption by LAr impurities or detector elements. This relatively long light path before a significant light quenching or absorption is achieved in ProtoDUNE-DP thanks to the excellent LAr purity and the absence of material inside the LAr active volume.

\begin{center}
\begin{table}[ht]
\caption{Attenuation length values of LAr scintillatoin light obtained from the exponential fits shown in Fig.~\ref{fig:RSL1}}
\centering
\begin{tabular}{@{}l*{7}{l}}
\br
 Approach & $\lambda_{att}$ (cm)\\
\mr
Data   &  180\,$\pm$\,17\\
MC (RSL 99.9\,cm)  &  180\,$\pm$\,10\\
MC (RSL 61.0\,cm)  &  157\,$\pm$\,8\\
\br
\end{tabular}\label{tab:RSL2}
\end{table}
\end{center}

\section*{Conlcusions}

ProtoDUNE-DP is a 6$\times$6$\times$6 m$^{3}$ LArTPC, operated at CERN in 2019-2020 to fully demonstrate the dual-phase technology for DUNE, the next generation long-baseline neutrino experiment. The photon detection system formed of 8-inch cryogenic PMTs collected cosmic-ray data in stable conditions. The slow scintillation constant, and therefore the LAr purity, was monitored during the whole data taking. ProtoDUNE-DP used PEN as wavelength shifter for the first time in a large scale experiment and a comparison with the widely used TPB is carried out. In ProtoDUNE-DP, it is obtained that, at least, 17\% of the scintillation light detected in absence of drift field comes from electron-ion recombination verifying the expected trend of the Birks’ law. The size of ProtoDUNE-DP, the longest drift-distance LArTPC ever operated, allows for an unprecedented study of the light propagation. The data are in agreement with a 99.9-cm Rayleigh scattering length. Cosmic muons were detected at a distance from the PMTs up to 6\,m.

\ack
This project has received funding from the European Union Horizon~2020 Research and Innovation programme under Grant Agreement no.~654168; from the Spanish Ministerio de Economia y Competitividad (SEIDI-MINECO) under Grant no.~FPA2016-77347-C2-1-P and MdM-2015-0509; from the Comunidad de Madrid; and the support of a fellowship from ''la Caixa" Foundation (ID 100010434) with code LCF/BQ/DI18/11660043.

\section*{References}
\bibliographystyle{iopart-num}  
\bibliography{biblio}

\end{document}